\documentclass[prl,twocolumn,showkeys,showpacs,superscriptaddress]{revtex4-1}
\usepackage{graphicx,latexsym,amssymb,amsmath,amsbsy,color}

\begin{document}

\title{Dissipation and Rheology of Sheared Soft-Core Frictionless Disks Below Jamming}

\author{Daniel V{\aa}gberg}
\affiliation{Department of Physics, Ume{\aa} University, 901 87 Ume{\aa}, Sweden}
\author{Peter Olsson}
\affiliation{Department of Physics, Ume{\aa} University, 901 87 Ume{\aa}, Sweden}
\author{S. Teitel}
\affiliation{Department of Physics and Astronomy, University of Rochester, Rochester, NY 14627}
\date{\today}

\begin{abstract}
We use numerical simulations to investigate the effect that different models of energy dissipation have on the rheology of soft-core frictionless disks, below jamming in two dimensions.  
We find that it is not necessarily the mass of the particles that determines whether a system has Bagnoldian or Newtonian rheology, but rather the presence or absence of large connected clusters of particles.  We demonstrate the key role that tangential dissipation plays in the formation of such clusters, and in several models find a transition from Bagnoldian to Newtonian rheology as the packing fraction $\phi$ is varied.  For each model we show that appropriately scaled rheology curves approach a well defined limit as the mass of the particles decreases and collisions become strongly inelastic.

\end{abstract}
\maketitle

Many seemingly disparate physical systems, such as granular materials, foams, emulsions, and suspensions, have been modeled in terms of soft-core  interacting particles in an athermal limit.  Such a common description has led to the prediction of common physical behaviors, notably the jamming transition from a liquid-like state, to a rigid but disordered solid \cite{LiuNagel1,LiuNagel2,vanHecke}.  Of particular interest has been the behavior of such systems under a steady shear strain rate $\dot\gamma$ \cite{OT,Hatano,Hohler,Boyer}. Below jamming, granular particles are usually described by Bagnoldian rheology \cite{Bagnold, daCruz, Lemaitre1, OtsukiHayakawa} with pressure $p$ and shear stress $\sigma$ scaling  $\propto \dot\gamma^2$ at low $\dot\gamma$.  However foams and emulsions are found to obey Newtonian rheology \cite{Boyer,Durian,Katgert,Cassar} with $p$, $\sigma\propto \dot\gamma$ at low $\dot\gamma$.  It is therefore important to understand what are the essential features of the microscopic interactions that lead to one rheology or the other.

Here we consider within a unified framework the effect that different, commonly used, models of energy dissipation have on the rheology of soft-core frictionless disks, below jamming in two dimensions (2D).  
We find that it is not necessarily the mass of the particles that determines whether a system has Bagnoldian or Newtonian rheology, but rather the absence or presence of large connected clusters of particles.  For Bagnoldian rheology we find, even in the dense limit, that as $\dot\gamma\to 0$, the average contact number $z\to 0$ and there are no instantaneous force chains.  In contrast, Newtonian rheology requires the formation of large connected clusters of particles, with extended force chains as jamming is approached.
We demonstrate the key role that tangential dissipation plays in the formation of such clusters, and in several models find a sharp transition from Bagnoldian to Newtonian rheology as the packing fraction $\phi$ is varied.  For each model we show that appropriately scaled rheology curves approach a well defined limit as the mass of the particles decreases and collisions become strongly inelastic.

Our soft-core model is as follows.  We take the elastic force on a particle at position ${\bf r}_i$ due to its contact with a particle at ${\bf r}_j$ to be
\begin{equation}
{\bf f}_{ij}^{\rm el} = -k_e\dfrac{dV(|{\bf r}_{ij}|/d_{ij})}{d{\bf r}_i},\quad {\bf r}_{ij}\equiv {\bf r}_i-{\bf r}_j.
\end{equation}
Here $d_{ij}=(d_i+d_j)/2$ is the average diameter of the two particles, $V(x)$ is a dimensionless soft-core interaction potential with $V(x)=0$ for $x>1$, and $k_e$ is the coupling that sets the energy scale of the interaction.  

For the dissipative force we consider several different models.  In the ``reservoir dissipation" model (RD) a particle with center of mass  velocity ${\bf v}_i=\dot{\bf r}_i$ decays to the average shear flow velocity. For uniform shear flow in the $x$-direction we have,
\begin{equation}
{\rm model\>RD}\qquad {\bf f}_i^{\rm dis}=-k_d [{\bf v}_i-\dot\gamma y_i{\bf \hat x}].\qquad\>
\label{erd}
\end{equation}
In the ``contact dissipation" model (CD) we assume dissipation is due to binary collisions between particles.  We take the force on particle $i$ due to contact with particle $j$ to be,
\begin{equation}
{\rm model\>CD}\qquad {\bf f}_{ij}^{\rm dis}=-k_d[{\bf v}_i-{\bf v}_j].\qquad\quad\>
\label{ecd}
\end{equation}
We also consider the model CD$_{\rm n}$ in which dissipation is due only to the velocity difference in the direction normal to the point of contact,
\begin{equation}
{\rm model\>CD_n}\qquad {\bf f}_{ij}^{\rm dis}=-k_d[({\bf v}_i-{\bf v}_j)\cdot{\bf\hat r}_{ij}]{\bf\hat r}_{ij}.
\label{ecdn}
\end{equation}
For theoretical completeness we also consider the model CD$_{\rm t}$,
\begin{equation}
{\rm model\>CD_t}\qquad {\bf f}_{ij}^{\rm dis}=-k_d[({\bf v}_i-{\bf v}_j)\cdot{\bf\hat t}_{ij}]{\bf\hat t}_{ij}.
\label{ecdt}
\end{equation}
where ${\bf\hat t}_{ij}\equiv {\bf\hat z}\times {\bf\hat r}_{ij}$ is tangent to the point of contact.

CD$_{\rm n}$ is typically used in models of massive dry granular particles \cite{Schafer}.  CD was introduced by Durian \cite{Durian} to describe the viscous interaction between massless foam bubbles. In that context, Durian also introduced RD as a mean-field approximation to CD \cite{Durian,Tewari}, in which the instantaneous velocity ${\bf v}_j$ is replaced by its ensemble average $\dot\gamma y_j{\bf\hat x}$.  However RD can also be considered as a model for  particles embedded in a uniformly sheared host medium, where dissipation is between the particles and the degrees of freedom that comprise the host, such as the Stokes drag on a particle in a fluid. Both CD and RD may be used to model massive particles in emulsions and suspensions \cite{Varnik}.

To model a uniform shear flow we use Lees-Edwards boundary conditions \cite{LeesEdwards}.  It is convenient to define the ``lab frame" coordinates ${\bf r}_i$ in terms of ``shear frame" coordinates $(x_i,y_i)$ that obey periodic boundary conditions, ${\bf r}_i\equiv (x_i+\gamma y_i, y_i)$, where $\gamma=\dot\gamma t$ is the total shear strain in time $t$.  
The equations of motion for a particle of mass $m_i$ can then be written as,
\begin{equation}
m_i(\ddot{x}_i+2\dot{\gamma}\dot{y}_i)=(f_{ix}-\gamma f_{iy}),\quad
m_i\ddot{y}_i=f_{iy},
\end{equation}
where ${\bf f}_i={\bf f}^{\rm el}_i+{\bf f}^{\rm dis}_i$ is the total force on the particle.  

Generalizing the work of Lema\^{\i}tre et al. \cite{Lemaitre2}, who considered only hard-core particles, we now cast our equation of motion into dimensionless form.  We take $d_s$ and $m_s$, the diameter and mass of our small particles, as our unit of length and mass, and $1/\dot\gamma$ as our unit of time.  Important time scales in the problem are the elastic and dissipative relaxation times,
\begin{equation}
\tau_e\equiv\sqrt{m_sd_s^2/k_e},\quad \tau_d\equiv m_s/k_d,
\end{equation}
as well as the time,
\begin{equation}
\tau_0\equiv\tau_e^2/\tau_d=k_dd_s^2/k_e
\end{equation}
which is independent of the mass scale $m_s$.
The degree of elasticity of collisions is conveniently expressed in terms of the ratio 
\begin{equation}
Q\equiv\tau_d/\tau_e=\sqrt{m_s k_e/(k_d^2d_s^2)}\propto \sqrt{m_s}.
\end{equation}
For a harmonic elastic interaction, a head-on collision will be totally inelastic (coefficient of restitution $e=0$)
when $Q<\frac{1}{2}[m_s d_{ij}^2/(\bar m d_s^2)]^{1/2}$, where $\bar m$ 
is the reduced mass of the colliding particles \cite{Schafer}.

The equations of motion for the dimensionless trajectories $\{X_i(\gamma),Y_i(\gamma)\}\equiv\{x_i(t)/d_s,y_i(t)/d_s\}$ are then, 
\begin{equation}
\begin{array}{rl}
\rho_i(\dot{\gamma}\tau_e)^2\left[X_i^{\prime\prime}+2Y_i^\prime \right]&= [F_{ix}^{\rm el}-\gamma F_{iy}^{\rm el}]+\dot\gamma\tau_0[F_{ix}^{\rm dis}-\gamma F_{iy}^{\rm dis}]\\[12 pt]
\rho_i(\dot{\gamma}\tau_e)^2Y_i^{\prime\prime}&=F_{iy}^{\rm el}+\dot\gamma\tau_0F_{iy}^{\rm dis},
\end{array}
\label{eom}
\end{equation}
where $\rho_i\equiv m_i/m_s$,  $X_i^\prime\equiv dX_i/d\gamma$. The dimensionless forces are,
\begin{equation}
\begin{array}{rl}
{\bf F}_{i}^{\rm el}(\{{\bf R}_i\})&\equiv(d_s/k_e){\bf f}_i^{\rm el}(\{{\bf r}_i\})\\[12 pt]
{\bf F}_i^{\rm dis}(\{{\bf R}_i^\prime\})&\equiv(1/k_d d_s\dot\gamma){\bf f}_i^{\rm dis}(\{{\bf v}_i\}),
\end{array}
\end{equation}
where the dimensionless lab frame trajectories ${\bf R}_i\equiv(X_i+\gamma Y_i,Y_i)$ depend on the strain rate only though the dimensionless strain parameters $\dot{\gamma}\tau_e$ and $\dot{\gamma}\tau_0=\dot\gamma\tau_e/Q$.


 
To study the rheology we are interested in the stress tensor.  Here we consider only the elastic part ${\bf p}^{\rm el}$, which dominates over the kinetic and dissipative parts for all but the largest $\dot\gamma$.  Since ${\bf p}^{\rm el}\equiv L^{-D}\sum_{i<j}{\bf f}^{\rm el}_{ij}{\bf r}_{ij}$ in $D$ dimensions \cite{OHern}, we can define a dimensionless stress tensor,
\begin{equation}
{\bf P}^{\rm el}\equiv \left(\dfrac{d_s}{L}\right)^D\sum_{i<j}{\bf F}^{\rm el}_{ij}{\bf R}_{ij}=\dfrac{\tau_e^2d_s^{D-2}}{m_s}{\bf p}^{\rm el},
\end{equation}
where ${\bf R}_{ij}\equiv {\bf R}_i-{\bf R}_j$.   ${\bf P}^{\rm el}$ thus depends only on the dimensionless trajectories ${\bf R}_i(\gamma)$, and so plotting the pressure $P^{\rm el}=\frac{1}{2}{\rm tr}[{\bf P}^{\rm el}]$ vs $\dot\gamma\tau_e$, all models with the same $Q$ will fall on the same curve, independent of the specific values of $k_e$, $k_d$, and $m_s$.  In particular, the hard-core limit $k_e\to\infty$, with $k_d\sim \sqrt{k_e}$ so that $Q$ stays constant, will also lie on the same curve.  Since $\tau_e\to 0$ in this limit, we conclude that the hard-core limit may be inferred from soft-core simulations, provided one looks at sufficiently small $\dot\gamma$. 

In our simulations of the above four dissipative models, we have observed two limiting forms of behavior: (i) the ``overdamped" limit, where the kinetic term is negligible and the trajectories ${\bf R}_i(\gamma)$ are determined by the balance of elastic and dissipative terms -- here one has Newtonian rheology at small $\dot\gamma$; and (ii) the ``inertial" limit, where the dissipative term is negligible and the trajectories are determined by the balance of elastic and kinetic terms -- here one has Bagnoldian rheology at small $\dot\gamma$.

(i) The overdamped limit results when the kinetic term in Eq.~(\ref{eom}) becomes negligible.  We then have,
\begin{equation}
{\bf F}_i^{\rm el}=-(\dot\gamma\tau_0){\bf F}_i^{\rm dis}.
\end{equation}
and the dimensionless trajectories thus depend parametrically only on the parameter $\dot\gamma\tau_0$. Assuming the trajectories have a well defined limit as $\dot\gamma\to 0$, then the leading dependence of ${\bf F}_i^{\rm el}$ on $\dot\gamma$, and so presumably the pairwise contact forces ${\bf F}_{ij}^{\rm el}$, and so also the stress tensor ${\bf P}^{\rm el}$, is $\propto\dot\gamma\tau_0$, and so one has Newtonian scaling.  
Deviations from Newtonian scaling will occur at larger $\dot\gamma$ due to soft-core effects; these become stronger and set in at lower $\dot\gamma$ as one approaches the jamming $\phi_J$. 
But the characteristic feature of the overdamped limit 
is that curves of $P^{\rm el}$, when plotted vs $\dot\gamma\tau_0$, will approach a common limiting curve for different $Q$.
The dimensionless $\tilde\eta_p\equiv P^{\rm el}/(\dot\gamma\tau_0)$ approaches a constant as $\dot\gamma\tau_0\to 0$, giving  the hard-core limit.  The pressure analog of viscosity is then $p/\dot\gamma=k_d\tilde\eta_p$.

(ii) The inertial limit results when the dissipative term becomes negligible.  Eq.~(\ref{eom}) then becomes,
\begin{equation}
\begin{array}{l}
F_{ix}^{\rm el}=\rho_i(\dot\gamma\tau_e)^2[X_i^{\prime\prime}(\gamma)+2Y_i^\prime(\gamma)+\gamma Y_i^{\prime\prime}(\gamma)]\\[12 pt]
F_{iy}^{\rm el}=\rho_i(\dot\gamma\tau_e)^2Y_i^{\prime\prime}(\gamma),
\end{array}
\end{equation}
and the dimensionless trajectories now depend parametrically only on the parameter $\dot\gamma\tau_e$.  
The leading dependence of ${\bf F}_i^{\rm el}$ on $\dot\gamma$, and so the stress tensor ${\bf P}^{\rm el}$, is $\propto (\dot\gamma\tau_e)^2$.  One thus has Bagnold scaling.
Deviations from Bagnold scaling will occur at larger $\dot\gamma$ due to soft-core effects; these become stronger and set in at lower $\dot\gamma$ as one approaches the jamming $\phi_J$. 
But the characteristic feature of the inertial limit is that curves of $P^{\rm el}$, when plotted vs $\dot\gamma\tau_e$, will approach a common limiting curve for different $Q$.
The dimensionless 
$B_p\equiv P^{\rm el}/(\dot\gamma\tau_e)^2$
approaches a constant as $\dot\gamma\tau_e\to 0$, giving the hard-core limit.  The Bagnold coefficient for pressure is then $p^{\rm el}/\dot\gamma^2=m_sB_p$.

\begin{figure}[h!]
\begin{center}
\includegraphics[width=3.5in]{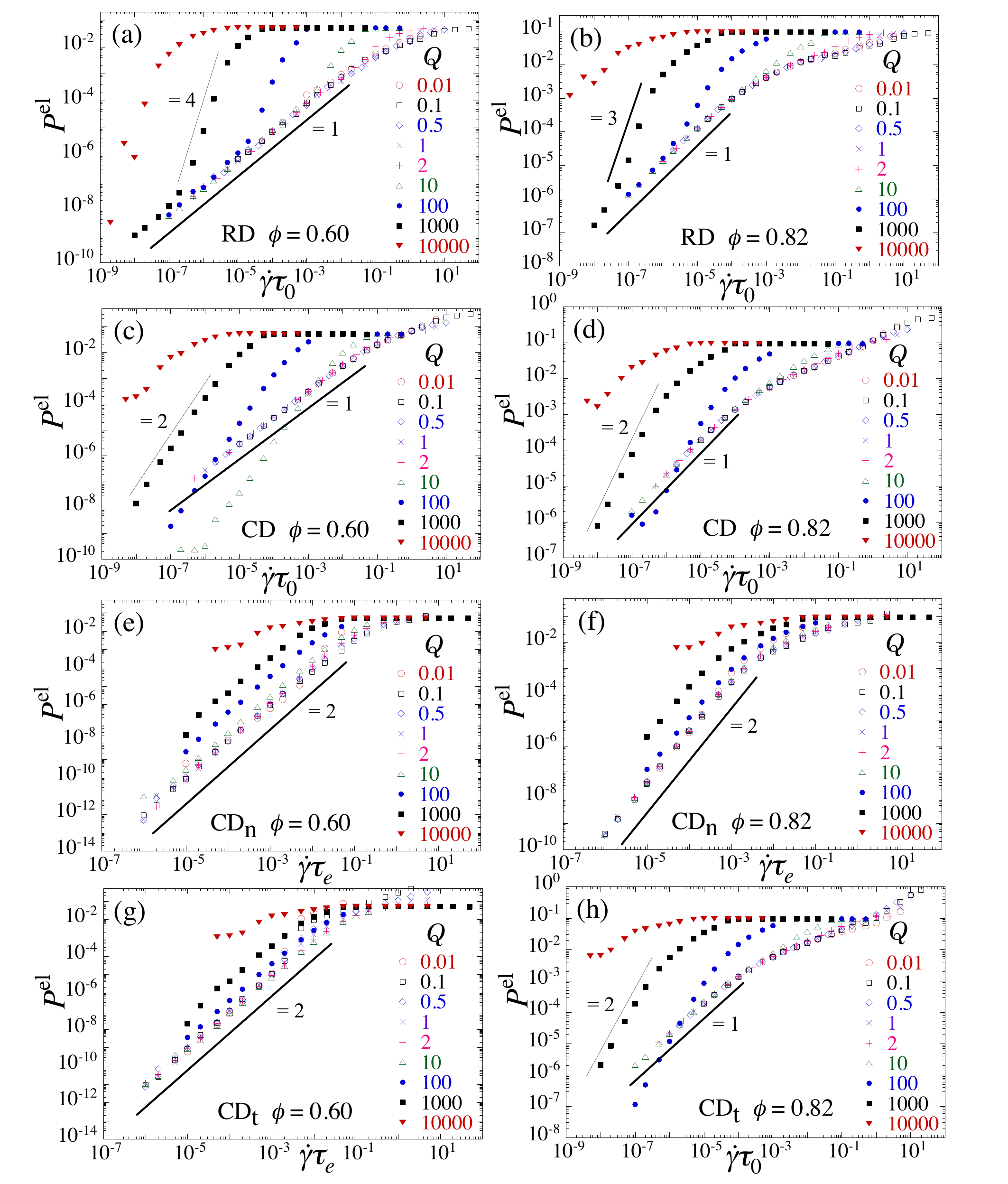}
\caption{(color online) Dimensionless pressure $P^{\rm el}$ vs dimensionless strain rate, $\dot\gamma\tau_0$ or $\dot\gamma\tau_e$, for different values of $Q$ for the four dissipative models of Eqs.~(\ref{erd}-\ref{ecdt}).  Left hand column is for packing fraction $\phi=0.60$; right hand is for $\phi=0.82$, close below jamming.  For each value of $Q$, several different choices of $m_s$ and $k_d$ were used.  Straight lines indicate algebraic behaviors, with power law as indicated by the neighboring number.
}
\label{f1}
\end{center}
\end{figure}

We now present the results of our numerical simulations.  Our simulations are for $N=1024$ total disks in 2D, with equal numbers of big and small particles with diameter ratio $d_b/d_s=1.4$, and $d_s=1$ \cite{OHern}.  Finite size effects are negligible for the range of parameters studied here.
Although the particles are of different size we take them to have equal mass $m_b=m_s\equiv m$. We simulate at fixed packing fraction $\phi = (\pi N/8L^2)(d_s^2+d_b^2)$, where the system area $L^2$ is varied to achieve the desired $\phi$.  We use a harmonic interaction, $V(x)=\frac{1}{2}(1-x)^2$ for $x<1$, with fixed elastic coupling $k_e=1$, and vary $k_d$ and $m_s$ to get different values of $Q$.  We integrate the equations of motion (\ref{eom}) using a modified velocity-Verlet algorithm with a Heun-like prestep to account for the velocity dependent acceleration.  We shear to a total strain $\gamma\sim 0.5 - 50$, depending on system parameters, collecting data only after the system appears to be in steady state.  Our main results for the four different dissipative models are presented in Fig.~\ref{f1}, where we plot the dimensionless pressure $P^{\rm el}$ vs the dimensionless strain rate, $\dot\gamma\tau_e$ or $\dot\gamma\tau_0$ (chosen according to the behavior we find in each particular model), for a wide range of $Q$.  We show results for a dilute case $\phi=0.60$, as well as for a dense case $\phi=0.82$ just below the jamming $\phi_J\approx 0.843$.

For model RD, Figs.~\ref{f1}a, b show Newtonian rheology at small $\dot\gamma$. At small $Q\lesssim 2$ we see the overdamped limit, with all data approaching a common limiting curve over 9 orders of magnitude in $\dot\gamma\tau_0$, spanning the range from Newtonian behavior at small $\dot\gamma\tau_0$ to non-Newtonian soft-core behavior, with accompanying shear thinning (slope $<1$), as $\dot\gamma\tau_0$ increases; as $\phi$ increases, the onset of this shear thinning moves to lower values of $\dot\gamma\tau_0$ as expected. 
For larger $Q$, the curves approach the common limiting curve as $\dot\gamma\tau_0\to 0$.
This is to be expected when the dimensionless ${\bf F}_i^\mathrm{dis}$ in Eq.~(\ref{eom}) is at least as big as the dimensionless kinetic factor ${\bf R}_i^{\prime\prime}$; then the kinetic term on the  left hand side of Eq.~(\ref{eom}) becomes negligible compared to the dissipative term on the right hand side, as $\dot\gamma\tau_0\to 0$ for any $Q$; one thus gets the overdamped limit.
This suggests that for RD the hard-core limit of $\tilde\eta_p$ is independent of $Q$ and hence of the mass $m_s$.  
However as $\dot\gamma\tau_0$ increases,
we see the onset of shear thickening (slope $>$1) due to inertial effects, as has been reported previously \cite{Andreotti}.  As $Q$ increases at fixed $\phi$, this shear thickening onset moves to lower $\dot\gamma\tau_0$; at fixed $Q$ it moves to lower $\dot\gamma\tau_0$ as $\phi$ increases.  
The saturation of $P^{\rm el}$ at large $\dot\gamma\tau_0$ represents the limit where particles have so much kinetic energy that the soft-core particles are able to pass through each other.   

For model CD, shown in Figs.~\ref{f1}c, d, behavior at low $Q$ appears qualitatively similar to that of RD; we are in the overdamped limit.  
In a separate work \cite{CDRD} we will argue that the criticality of the jamming transition for CD is the same as for RD as $Q\to 0$.  However, as $Q$ increases we see a transition at $Q^*$ from Newtonian ($\sim\dot\gamma$) to Bagnoldian ($\sim\dot\gamma^2$) rheology \cite{Fall}.  Comparing Fig.~\ref{f1}c with d, we see that $Q^*$ increases with increasing $\phi$. 

For model CD$_{\rm n}$, Figs.~\ref{f1}e, f show Bagnold rheology at small $\dot\gamma$ for all values of $Q$.  At low $\dot\gamma\tau_e$ we see shear thickening, with a slope $\sim 2> 1$, but as $\dot\gamma\tau_e$ increases, we see a crossover to shear thinning (slope $<1$) due to soft-core effects.  As $\phi$ increases, this departure from Bagnold rheology moves to lower values of $\dot\gamma\tau_e$ as expected.  In both Figs.~\ref{f1}e, f, we see that the inertial limit holds, with the data approaching a common limiting curve over 7 orders of magnitude in $\dot\gamma\tau_e$, for a range of small $Q\lesssim 2$, extending to larger $Q$ as $\phi$ increases. 
However, while our smallest $Q=0.01$ agrees with this limiting curve at small $\dot\gamma\tau_e$, it shows a clear departure increasing towards larger values of $P^{\rm el}$ as $\dot\gamma\tau_e$ increases.  

Finally, our results for CD$_{\rm t}$ are shown in Figs.~\ref{f1}g, h.  Here we find Bagnoldian rheology and the inertial limit, similar to model CD$_{\rm n}$, at the lower $\phi=0.60$.  However we find Newtonian rheology and the overdamped limit at the denser $\phi=0.82$, where behavior becomes very similar to that of CD. 
Thus, in contrast to CD, where we only find a transition from Newtonian to Bagnoldian rheology at large $Q$ where inertial effects become important, for CD$_{\rm t}$ we see such a transition as $\phi$ increases even in the limit of $Q\to 0$, i.e. $m_s\to 0$.

\begin{figure}[h!]
\begin{center}
\includegraphics[width=3.5in]{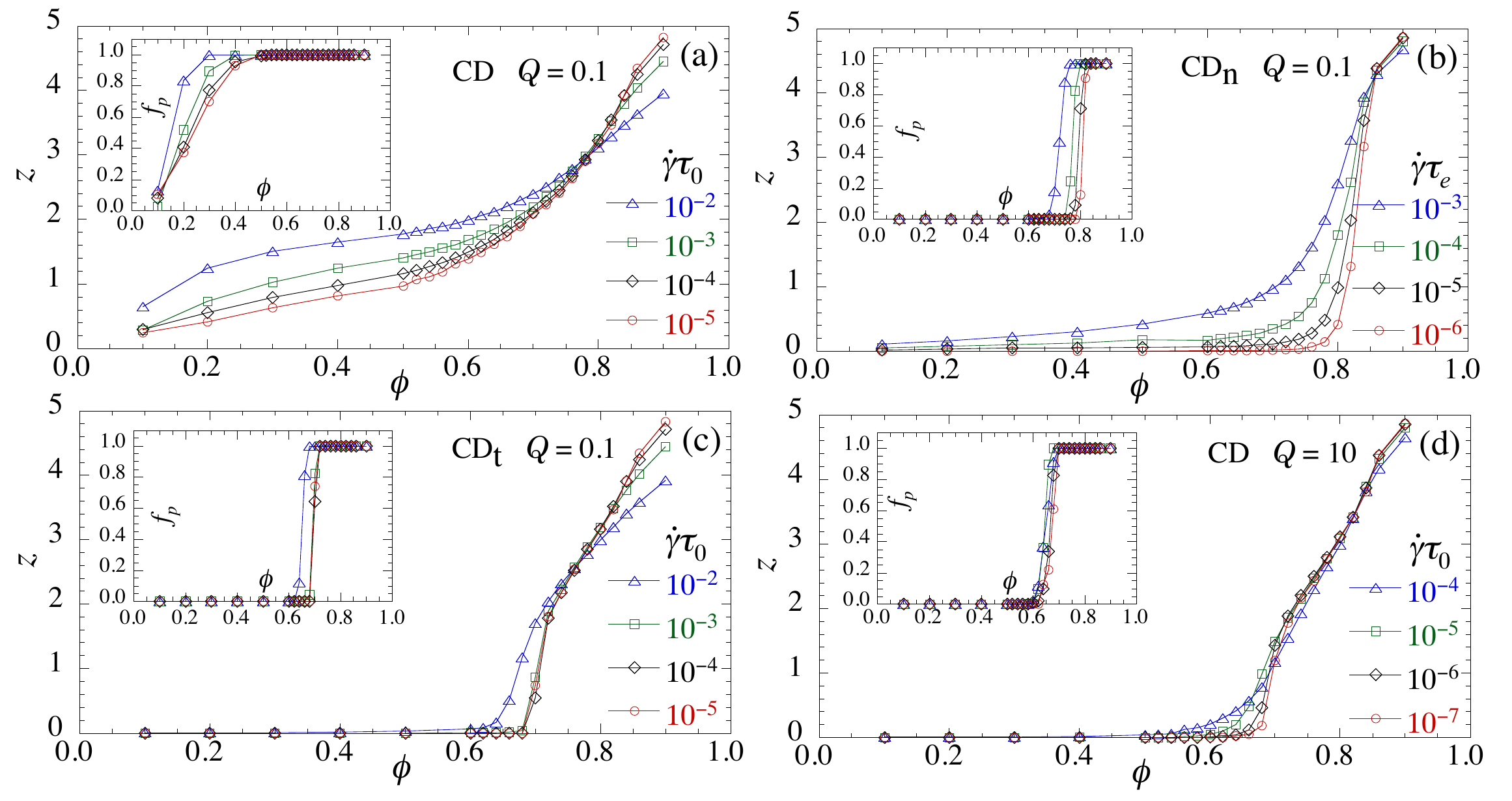}
\caption{(color online) Average contact number $z$ vs $\phi$ at different strain rates $\dot\gamma$ for models (a) CD, (b) CD$_{\rm n}$, and (c) CD$_{\rm t}$ at $Q=0.1$, and (d) CD at $Q=10$.  Insets show the fraction of states $f_p$ with percolating connected clusters.
}
\label{f2}
\end{center}
\end{figure}

To help understand the origin of the different rheologies, we note that for the collisional models CD, CD$_{\rm n}$, CD$_{\rm t}$, the overdamped limit is associated with the formation of large clusters of particles (for RD see \cite{supplement}).  When the majority of particles cohere together into clusters, particle accelerations become negligible and hence the kinetic term in the equation of motion can be neglected.
This connection is shown in Fig.~\ref{f2} where we plot the average particle contact number $z$ vs $\phi$, for different values of the applied strain rate $\dot\gamma$.  The inset to each figure shows the fraction of states $f_p$ which contain a percolating connected cluster of particles \cite{OHern2}. 

Figs.~\ref{f2}a-c are for strongly inelastic collisions, $Q=0.1$.
In Fig.~\ref{f2}a for model CD, where the rheology is overdamped, $z$ stays finite down to low $\phi$; the percolation fraction remains $f_p>0$ down to similarly low $\phi$.  The reason for this is simple. For such strongly inelastic collisions, the velocity difference 
of two colliding particles decays to zero during the collision, and the particles remain in contact. In Fig.~\ref{f2}b for model CD$_{\rm n}$, however, where the rheology is in the inertial limit, $z$ and $f_p$ drop rapidly to zero as $\phi$ decreases below $\phi_J\approx 0.843$; the drop sharpens as $\dot\gamma$ decreases, suggesting that $z,f_p\to 0$ for all $\phi<\phi_J$ as $\dot\gamma\to 0$.  Again the reason is simple.  Although the normal component of the velocity difference 
decays to zero during a collision, the tangential component
remains finite and causes the particles to move apart, breaking contact. In Fig.~\ref{f2}c we show model CD$_{\rm t}$.  Here we see that $z$ and $f_p$ remain finite as $\phi$ decreases below $\phi_J$, but they drop sharply to zero at $\phi^*\approx 0.7$; this marks the transition from the inertial limit at $\phi<\phi^*$ to the overdamped limit at $\phi>\phi^*$, as seen in Figs.~\ref{f1}g, h.  
Finally in Fig.~\ref{f2}d we show model CD again, but now for the case of large $Q=10$ where inertial effects are important.  We see a transition, with $z$ and $f_p$ dropping sharply to zero just below $\phi\approx 0.7$, marking the transition from Bagnoldian rheology at low $\phi$ to Newtonian rheology at high $\phi$, as seen in Figs.~\ref{f1}c, d.  

We can qualitatively explain the observed transitions as follows.  When the system is dilute, particles separate whenever the total velocity difference of the colliding particles is not damped to zero during the collision.  This occurs for model CD$_{\rm t}$ (CD$_{\rm n}$) even in the strongly inelastic limit of small $Q$, since the normal (tangential) component of the velocity difference does not get damped at all.  For CD it happens only at larger $Q$ when collisions are less inelastic.  For dense systems, however, many body effects become important. 
At sufficiently dense $\phi$, normally directed relative particle motion becomes energetically prohibitive; it is a compressive motion that would induce particle overlaps, and so is constrained by the dense particle geometry.  This is in contrast to tangential relative particle motion which corresponds to a local shear deformation with particles sliding around each other with minimal overlaps.  Thus in CD$_{\rm n}$, where tangential relative motion is not damped, particles continue to separate after collisions.  But in CD$_{\rm t}$ and CD, where tangential relative motion is damped, particles form clusters. Indeed we find that for all models, at densities $0.6\lesssim\phi$ the relative motion of particles in contact is almost always tangential, with
the particles' separation ${\bf r}_i-{\bf r}_j$ very nearly orthogonal to their relative motion ${\bf v}_i-{\bf v}_j$ \cite{supplement}.
A similar result was found for the response of statically jammed packings to a small shear deformation \cite{Ellenbroek}.

To conclude, we have shown that the rheology of soft-core frictionless disks is strongly dependent on the specific form of the dissipative interaction.  At dense $\phi$ in collisional models, tangential dissipation is crucial for the particle clustering that gives Newtonian rheology.  Bagnoldian rheology results when particles separate after collisions and the average contact number $z\to 0$.  Sharp transitions between Bagnoldian and Newtonian rheology may exist as a function of particle density $\phi$ and the degree of inelasticity of collisions as measured by $Q$.
In the small $Q$ (small $m_s$) regime of strongly inelastic collisions, the rheology curves approach a limiting form in both the overdamped and inertial cases, that extends from the low $\dot\gamma$ hard-core limit into the higher $\dot\gamma$ region where soft-core effects are manifest.

\section*{Acknowledgements}

This work was supported by NSF grant DMR-1205800 and Swedish Research Council grant 2010-3725. Simulations were performed on resources
provided by the Swedish National Infrastructure for Computing (SNIC) at PDC and HPC2N.  We thank B. Tighe, H. Hayakawa, and C. Maloney for helpful discussions.

\section*{Supplemental Material}
\setcounter{figure}{0}

\subsection{Relative Motion of Contacts}

\begin{figure}[h!]
\begin{center}
\includegraphics[width=3.5in]{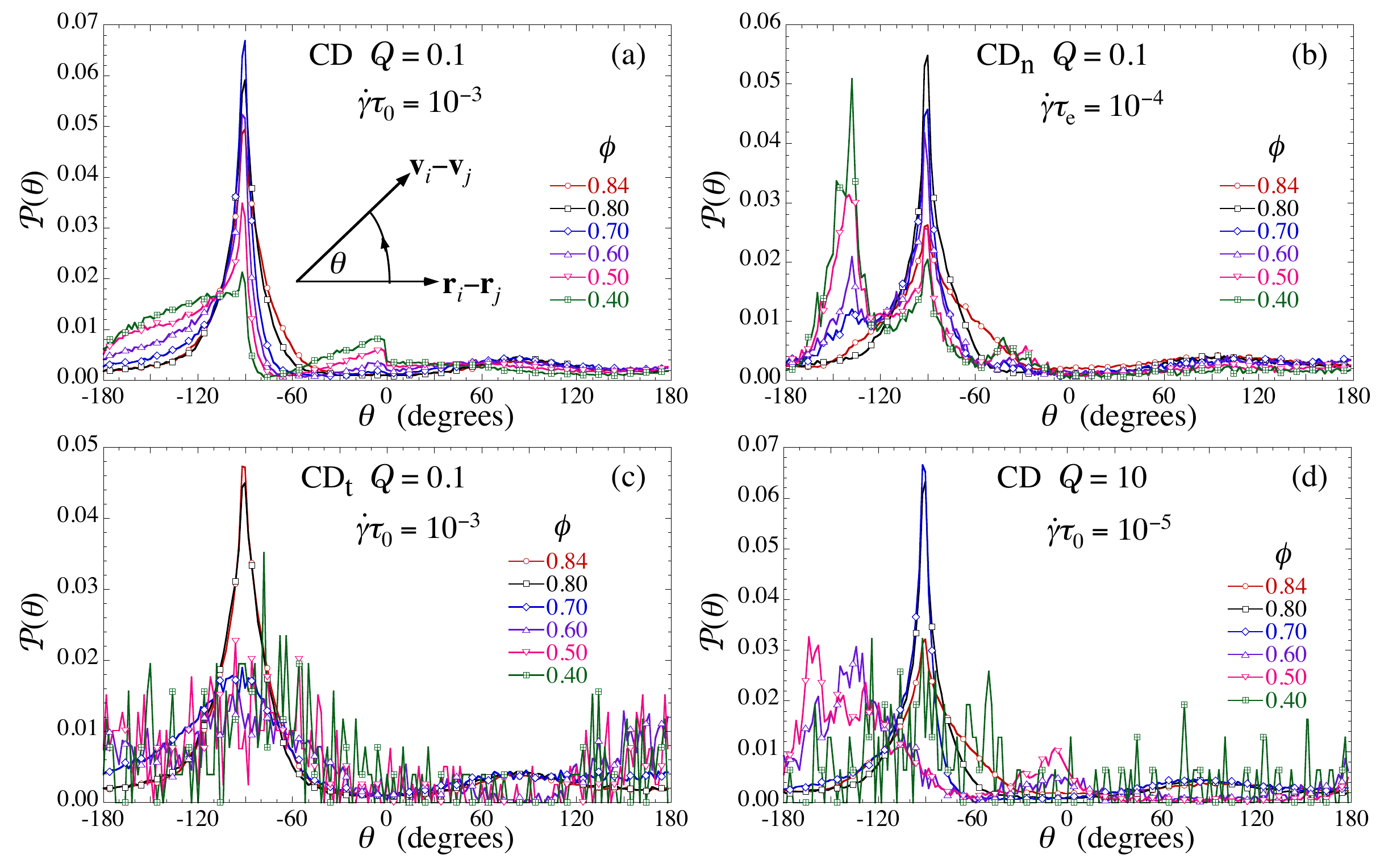}
\caption{(color online) Histograms of the angle of contact $\theta$ for (a) CD, (b) CD$_{\rm n}$, and (c) CD$_{\rm t}$ at $Q=0.1$, and (d) CD at $Q=10$.  Only every fifth symbol is plotted for clarity.  Inset to (a) shows the definition of $\theta$.
}
\label{f1SM}
\end{center}
\end{figure}

To characterize the nature of the shear-induced particle collisions  in our soft-core models, we consider a quantity that we call the angle of contact $\theta$.  We define this as the angle that the velocity difference ${\bf v}_i-{\bf v}_j$ makes with respect to the particle separation ${\bf r}_i-{\bf r}_j$ for two particles in contact.  
In Fig.~\ref{f1SM} we show plots of the histogram ${\cal P}(\theta)$ of the angle of contact $\theta$ for our different collisional models.  Figs.~\ref{f1SM}a, b, c  are for models CD, CD$_\mathrm{n}$ and CD$_\mathrm{t}$ respectively, in the strongly inelastic case of $Q=0.1$.  Fig.~\ref{f1SM}d is for model CD in the weakly inelastic case of $Q=10$.  We see that for all models, for the denser values of $\phi\gtrsim 0.6$, ${\cal P}(\theta)$ shows a strong peak at $\theta=-90^\circ$, i.e. we have primarily tangential relative motion at contacts.  There is essentially  no normal relative motion at $\theta=0$, except at low $\phi$. 

\subsection{Rheological Curves}

In Fig.~1 of the main paper we presented plots of the dimensionless elastic part of the pressure $P^\mathrm{el}$ vs an approriate dimensionless strain rate $\dot\gamma\tau_0$ or $\dot\gamma\tau_e$.  The choice of $\tau_0$ was used for systems with overdamped Newtonian rheology at small $Q$, while $\tau_e$ was used for systems with inertial Bagnoldian rheology at small $Q$, so that the data for small $Q$ collapses to a common curve in each case.  It is interesting to look at such rheology curves but now plotted with the opposite choice for dimensionless strain rate, i.e. where in Fig.~1 of the main paper we had plotted vs $\dot\gamma\tau_0$, here we plot vs $\dot\gamma\tau_e$, and vice versa.  We show such plots in Figs.~\ref{f2SM} below.
As should be expected, we now no longer see any simplifying data collapses at any values of $Q$.

\begin{figure}[h!]
\begin{center}
\includegraphics[width=3.5in]{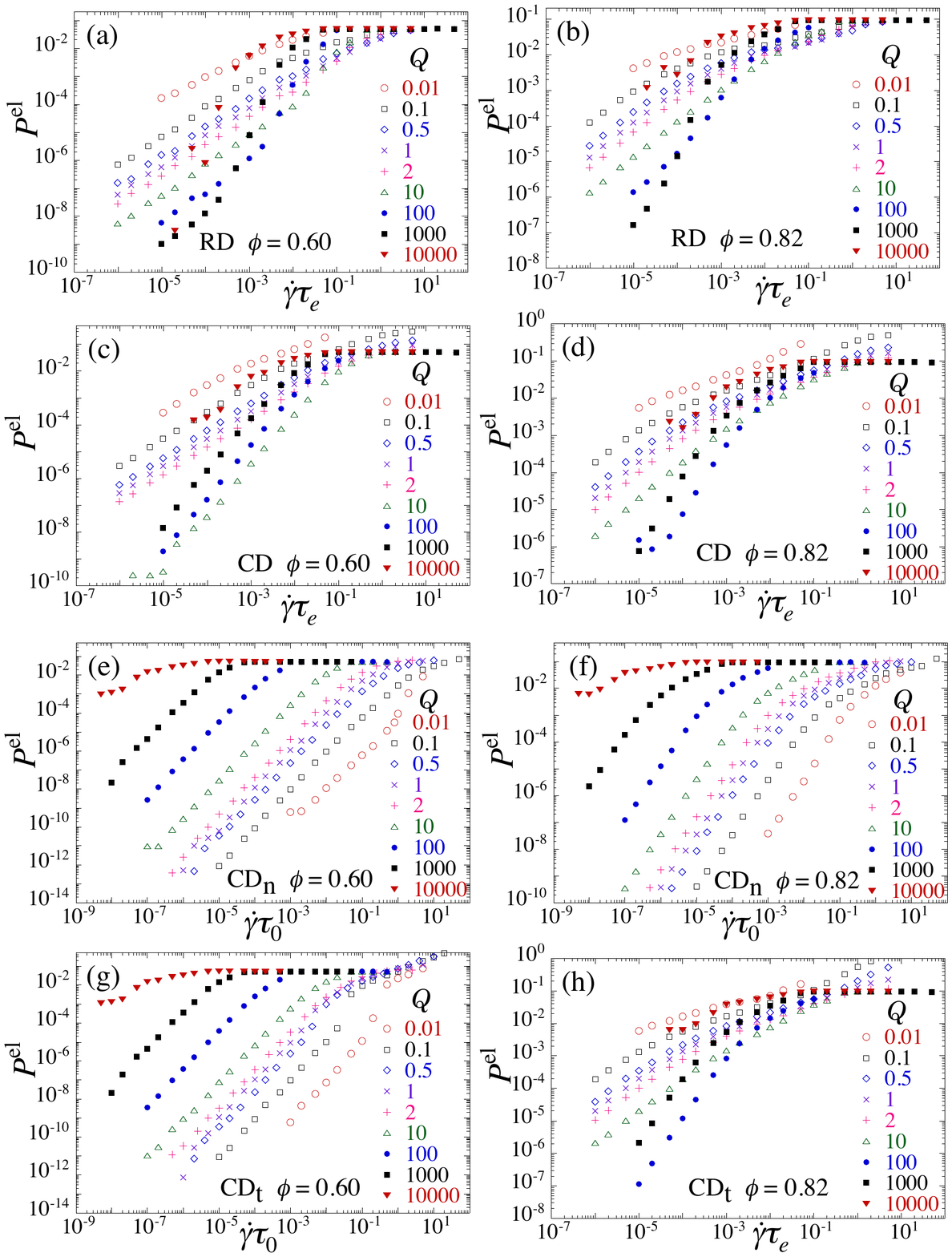}
\caption{(color online) Dimensionless pressure $P^{\rm el}$ vs dimensionless strain rate, $\dot\gamma\tau_0$ or $\dot\gamma\tau_e$, for different values of $Q$ for the four dissipative models defined in the main paper.  Left hand column is for packing fraction $\phi=0.60$; right hand is for $\phi=0.82$, close below jamming.  For each value of $Q$, several different choices of $m_s$ and $k_d$ were used.  
}
\label{f2SM}
\end{center}
\end{figure} 

\subsection{Particle Clustering in Model RD}

Our discussion of the relation between rheology and clustering, as shown in Fig.~2 of the main article, was limited to the collisional dissipation CD-models.  Here we discuss the situation for the reservoir dissipation model RD.

Newtonian rheology results whenever the dissipative term dominates over the kinetic term.  For the CD-models, where energy dissipation is due to binary particle collisions, the strength of the dissipative term depends on how long a given collision lasts (collision time is short when particles separate after colliding, collision time is long when particles stick together after colliding).  For RD, a particle's energy dissipation is with respect to the uniform sheared background, with which the particle is {\em always} in contact.  Hence the dissipative term never becomes negligible and we always have Newtonian rheology at small $\dot\gamma$.  

The presence of Newtonian rheology in RD is thus not necessarily related to particle clustering as it is for the CD-models.  Nevertheless we can still ask how the average contact number $z$ and the percolation probability $f_p$ vary with $\phi$ for model RD.  We show these quantities in Fig.~\ref{f3SM} below, for several different strain rates $\dot\gamma$ for the overdamped case of $Q=0.1$.  We see that the contact number $z$ stays finite for all $\phi$, with no strong dependence on $\dot\gamma$.  Thus particles tend to remain in contact with other particles, unlike the case when one has Bagnold scaling where $z\to 0$ as $\dot\gamma\to 0$.  The percolation probability stays roughly equal to unity above $\phi\approx 0.6$, but then drops rapidly to zero below.  Thus at low $\phi$ the particles are in clusters, but the clusters do not percolate across the system.  We do not yet understand if this percolation transition in RD at $\phi\approx 0.6$ has any physical consequences; it does not appear to effect the rheology.  

\begin{figure}[h!]
\begin{center}
\includegraphics[width=2.5in]{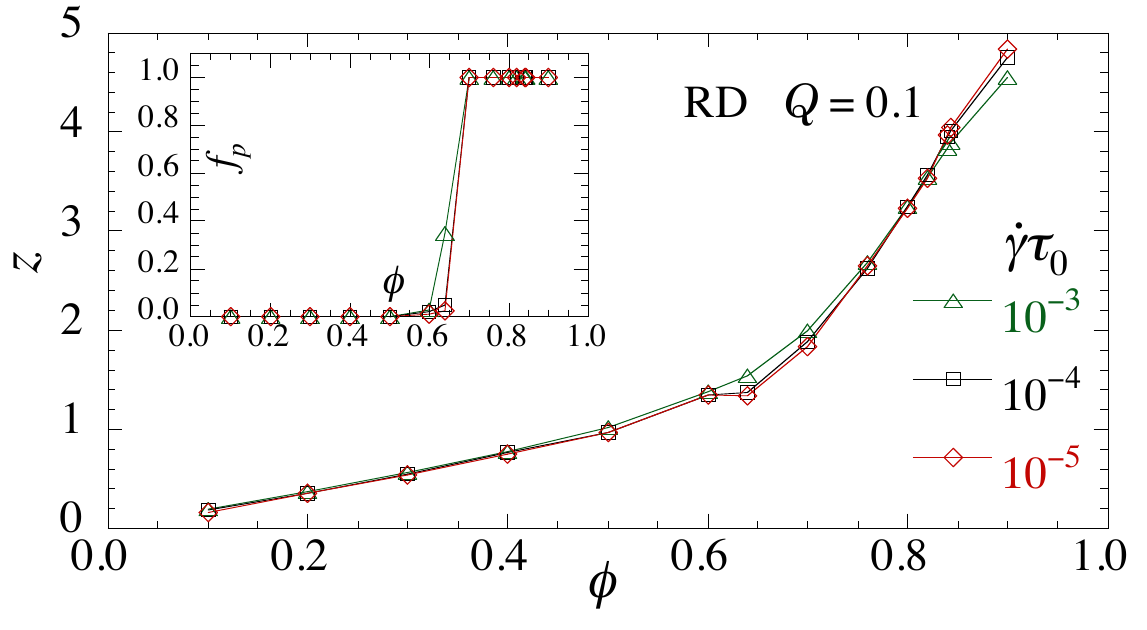}
\caption{(color online) Average contact number $z$ vs $\phi$ at different strain rates $\dot\gamma$ for model RD in the overdamped limit, $Q=0.1$.  The inset shows the fraction of states $f_p$ with percolating connected clusters.
}
\label{f3SM}
\end{center}
\end{figure}

\end{document}